# Investigation of functional systems of the psychic self-organization based on the method of basal matrix

V.T. Volov, V.V. Volov (j)

Abstract

The innovational matrix method of the basal emotions analysis for individuals is presented. The matrix criteria of the psycho-emotional state stability have been formulated. The developed method approbation on the myographic research data for individuals and comparison with psychological estimations results have been carried out.

Among the fundamental questions of scientific psychology the problem of the psychic stability estimation is especially important. This indicator is the key for the psychic state determination and forecast.

Zalevsky`s investigations were devoted to excess stability – rigidity [4]. On the long time investigation basis Zalevsky has formulated the fixed behavior forms theory, which opens the basis of deviations on the position of the psycho-emotional stability analysis of the individual. The researches of L.A. Kitaev-Smik, A.S. Bobrovoy, A.O. Prohorov, A. G. Dikoy were devoted to the psychic stability problem too.

In these investigations this phenomenon is researched in traditional aspect: as a manifestation of self-regulation due to the formation of the necessary state. With this regard, the emphasis is placed on revealing the personal, charactiriological and emotional characteristics of an individual.

Besides conscious regulation by the person of own psycho-emotional state, it has place natural often extramental mechanisms which consist in the mental live basis. In psychology this sphere has been determined as subconsciousness the manifestation of the one is discoved in so-called protected mechanisms, conversion syndrome etc.

Nowadays the methods and methodological apparatus of this phenomenon are absent. Among all variety of the psycho-diagnostic methods we cannot meet the one solves a problem of the psycho-emotional state measuring with the subsequent criteria estimation of stability at this point in time.

Existing test devices offer the possibility of estimation for the neuro-psychological characteristics, the types of nervous system, temperature, etc.

Tests have been developed for the actual psycho-emotional state assessments which are based on the self-reporting of subject or diagnostic signs of mental deviations i.e. offering qualitative assessments.

With the solution of the stability psyche as a system, investigation the study of the emotion`s sphere is connected.

As known, emotions present themselves complex psycho-physiological phenomena accompanying without exception all mental processes, which differ in the complexity of their organization, solving various tasks associated with self-organization, communication and adaption.

Every form of the emotional regulation is accompanying the certain physiological changes.

The definition of sustainability of any system starts with quantitative assessments.

Concerning the psycho-emotional stability the given task, in our view can be adequately solved by natural science methods, with using the psycho-physiological measurements. Among them the biggest interest has presented the investigation of emotions on the electromyografic measurements (EMG) of face (the) reaction basis. In accordance with the hypothesis of facial feedback, mimic reactions along with other physiological changes not only presides emotion in the form of bodily sensations but also constitutes its essence (S. Tomkins, K. Izard, A. Gelgorn) [3].

In our investigation we proceed from the idea of the facial feedback as a physiological manifestation of a feedback mechanism – the main link in the emotions regulation system in which facial reactions capture and encode a certain state (P.K. Anokhin).

Besides, we are based on the energy paradigm in the psyche stability assessment on the emotional phenomenon investigation that corresponds to foundations of the information and consumer theory (V.S. Simonov). A landmark in the psychic stability investigation of emotions was a definition of these existences and necessary measurements for their assessments with the help of which one can set the level of balance and tension within the system – the mimic reactions. The ability to measure given characteristics in the feedback actualization regime provides unique conditions for the emotion investigation as the functional system.

Besides, the research of emotions in given paradigm allows to clarify such phenomenon as the emotional resonance effect, empathy, phenomenon of alexithymia, etc.

An experimental design defined a subject area of this research – an organization of functional systems of psychic self-organization in extreme conditions of the central nervous system (CNS). Mimic people's reactions of patients with epilepsy as mechanisms for achieving and maintaining mental stability due to percularities of emotional response is the most manifested in conditions of the paroxysmal brain are investigated in the work.

The mimic investigation is carried out in traditional aspect of notorious positions of the nonverbal reactions in the various emotional states.

Unfortunetly, inspite of the rich palette the same works (A. Piz, P. Akman, U. Phrisen, K. Izard i.e), which connects with detailed detection of the psycho-motor regulaties for the emotional manifestation.

The researches of mental stability in scientific and practical relation at such approach are left in the basket.

Due to the above said, the purpose of this research is development of the quantitative stability criteria and qualitative signs of the psycho-emotional sphere (PES) state determination.

On the methodological basis of the investigation we use the following methods: methods of the clinical analysis, supervision, facial myography, mathematic methods (matrix approach, entropy and statistic analysis).

## 1. Qualitative-quantitative method of the psycho-emotional state estimation for individuals

In the presented monograph investigation 40 patients with epilepsy and 20 healthy mental balanced people were involved. Measurements for each individual were realized separately.

The averaged, confidence interval of the MBE elements, built on the myographic research basis equals $\Delta n = 3.57 \; mW$.

Originally, the myography measurements of «mask» were realized i.e. the state of tension of facial muscles before investigations. Then it was carried out the definition of mimic scheme at the experimental probes. At the first probe the patient was offered to imagine the situation in which he felt a certain emotional state. With it the task was put to feel this experience the maximum possible.

In the following sample it was shown a series of photographs of faces which partially showed basal emotions: anger, fear, joy, surprise (test P. Akman).

During the short-term view of stimulus material (1,5-2 s.) the patient had to identify the emotion. An additional procedure was a sample on an image of emotions. A list of the same basal emotions, which need alternately to turn to portray without emotional experience is suggested. Experiment duration of emotions was 5 seconds a pause between samples for each emotion was 10 seconds.

During the execution of the samples a measurement of the tension of the myography facial zones, connected with mimic patterns of emotions were conducted.

The assessment of the facial reactions quality produced on the maximum amplitude of muscle contraction: increase or decrease the level of «mask» tone were considered, respectively.

Primary analysis of mimic reactions in the first sample revealed the pattern deformations of basal emotions which are characteristic to patients with epilepsy.

Identified the following distortions:

1) reactions are completely inappropriated to the standard;

2) partially distorted patterns;

3) antagonistic patterns.

From the qualitative analysis it must follow that in the main group the mimic pattern disturbances of basal emotions are met more often, which are manifested in the contrary of standard reactions or in their absence in separate leads. Paradoxal reactions in the form of the amplitude and frequency contractions decreasing have been identified on the feelings of fear and sadness with leads 1 $(C\uparrow)$ and 4 $(Z\downarrow)$ correspondingly. In the arbitrary group the same tendations aren`t manifested. Frequently the total type of reaction is met which is manifested in general increase of the muscle tone on all leads independently from the emotional pattern. For the healthy people group this tendation doesn`t take place. Deviation of the pattern immediately on three leads with the opposite direction inherent for epileptics characterizes the so-called antagonistic reactions.

These reactions show us not random decay of the emotion but more often its block.

Antagonistic reactions such as other manifesting of the basal affect block of epileptics are watched often and predominantly are realized on emotions of fear, sadness and anger.

The total block can be considered as a general manifestation of an antagonistic block on the emotion of fear. In healthy people group the block manifestations are met less frequently: more often this is not a complete pattern of emotion.

For example, on one or two leads are marked corresponding reactions but on the one is absent. In this case with antagonistic reactions they are identified as on common with a general group emotions of fear and sadness so and on joy. This phenomenon is not watched in the patients with epilepsy. Besides in epileptic group not rarely zero reactions immediately on two (or more) leads are showing the reduced emotional response and the degradation of the facial feedback.

The second sample also identifies differences as on quality of emotion identification so and on myographic indicators. The first probe stage connects with definition of emotions. The control group identifies the emotions significantly better and on separate points-leads are fixed corresponding to the standard pattern.

This fact shows the right work out feedback as on the afferent sinthes so on the efferent level of the emotional system regulation (ESR) link.

An emotion of joy is well recognized in two groups. In general group is shown not only more wrong answers but it is false emotion recognition when given in the stimul data of the faces with manifestation of one emotion is stable taken for another. It is interesting to emphasize that itself false determined emotion is not identified during presentation of the one to epileptics. Sometimes patients identify indifference and other emotions which don`t include in the stimul data.

The false emotional identifying of surprise for fear and reverse is typical for both groups.

However, the trend of the one is not specific here it is appropriate to all patients (Akman`s data). Among emotions fear is identified by epileptics the worst. Given tendention is occurred not random: the myographyc data of the second group have identified the certain regularities. Aspecialy often it is observed on fear and sadness.

In the second stage of the probe by using EMG was measured facial reactions in parallel with the first stage. In the main group under perception of emotions revealed paradoxical reactions of the facial muscle tone reducing instead of tension (or reverse) and the absence of changes, as in the first sample. This tendention is observed in several leads: on fear, sadness and angry. The similar reactions are met in the control group, however in contrast to the general group it only has place on wrong identified emotions and much more seldom.

In the main group not corresponding to the standard reactions are met not only more frequently, but under mistaken recognition and with the correct perception of emotions. The last type of response indicates the limitation of the facial feedback and not about the pattern decay of emotions. The decay of emotions characterizes these profiles of the samples in which the broken pattern is random, i.e. does not repeat the pattern of the first test and due to a violation of a feedback mechanism of SER.

Facial deviations of the mimic pattern from the sample in the second sample in the identification of emotions characterizes this trend as isolation of an affect – is a special form of block. This can be partially broken pattern, or an antagonistic pattern. In the group of patients with epilepsy is identified the opposite trend when mispriced emotions showed compliance with the standard in the second sample when the properly mimic pattern is in the first sample. Among the rare trends of epilepsy patients it is met reactions with the correct pattern in partial recognition of emotions, when in the first sample pattern is violated. These reactions indicate that the emotion is not available for experiencing manifests feedback independently from the operation of this mechanism at the level of reflection. Thus, the described non-standard trends identified through fear, sadness and anger, demonstrate variations of the isolation mechanism of emotions of patients with epilepsy, which are manifested in a partial blocking of emotions at the level of unconscious or cognitive perceptual channels of the afferent facial feedback link.

As a result of the comparative analysis of data of two samples from patients with epilepsy revealed similar abnormalities in the form of mirror reactions, which are repeated in the experimental situation record of emotions and in situations of the mimic reflection during the perception of emotions. So, affect with the antagonistic reaction in the first sample is often observed in the second sample, even when there is no proper definition. It is met mirrored reactions of the disturbed pattern in the form of a playback of a total block of the first test to the second one. Also, there is a recurring profile of reactions with the partially disrupted pattern of the basal emotion. Thus the phenomenon of blockade of emotions is manifested due to the limitation of the efferent link of the feedback mechanism of SER. Here the violation of the emotion pattern is not a sign of decay. The repetition of antagonistic reactions on the same emotions in the group of healthy subjects does not occur, as other patterns.

On the basis of comparisons of the nature of these reactions of two samples and of the data analysis on identification these of emotions have shown tendentions, which are defined as the closure and isolation of emotions. Both mechanisms are associated with specific changes in the pattern of emotion which are revealed in the samples with the model reaction accounting and the possibility their determination. The block is shown in following reactions:

1) an antagonistic reaction in the first sample;

2) any chiralic trends with abnormal pattern (including of an antagonistic block) in the recognition of emotions in the second sample;

3) asymmetric trends (a special form of the chiralic effect), regardless of recognition;

4) the total block.

In the case of the emotion block it explaines the limit at the level of the facial expression – efferent link of system response, both in identification at the time of its perception, and without. The phenomenon of the block is in the emotion limiting and is associated with maintaining static stability of PES. Accounting the afferent link work, diagnosed in the identification of emotions and when compared with the reactions of both samples, it allows to determine the nature of the block. For example, a block is defined when there is a broken pattern in both samples with the chiralic effect under identifying of the emotion. Despite the availability for the perception of emotions, it does not manifest itself in the expression of the face properly, and, therefore, the development of the state, due to itself doesn't occur.

Diagnosis of reactions of the facial feedback in the second sample in the experiment is a model of monitoring of the afferent link of feedback of emotions as a functional system. This approach allows to determine not only the nature of the SER violation, but also to identify the links of the functional systems involved in compensatory mechanisms associated with paroxysmal conditions. On the basis of determining of the emotion identity in the experiment, as well as chiralic effects in the samples (showing the operation of identifying at the level of unconscious perception), is the work of an afferent link of emotions. Typical limitations of this link in comparison with the quality of the patterns of two samples allow to determine the isolation mechanism.

The following reactions are correlated to the isolation mechanism:

1) an antagonistic reaction in the second sample;

2) chiralic trends with a partially disrurbed pattern in the absence of identification of emotions;

3) the violation of the pattern of emotions in the second test in the recognition (sometimes with the right pattern in the first sample);

4) falsely recognizable emotions.

The case of the SER feedback limitation in the first paragraphs has been reflected. In particular, the second type of reaction is the removal of the function of awareness of emotions in the process of work out of the emotional resonance effect – is a cognitive channel of afferent synthesis of SER. In the third type of regulation, on the contrary, the channel of afferent synthesis is triggered at a cognitive level, but at the level of the efferent synthesis is the restriction of emotions in the form of the similar distorted pattern. The fourth type of response is the sole way of isolation of emotions: the perceptual channel of afferent synthesis is triggered under fixing of the cognitive channel.

In additional sample on the emotional image besides the collapse of facial patterns in the main group other differences are identified. Mainly the measurements were performed by an average amplitude and frequency of contractions. The group of patients with epilepsy is characterized by super-high tension of facial muscles (8-10 times), mainly for diversion No. 4. Moreover, the tension on the three leads is a disproportionate nature. This distinguishes a core group from the healthy persons where the tension is higher than in the first two samples in 3-5 times. In our opinion, these differences reveal the SER violation, and its instability.

**The generalization of the results.** In the main group on the first two samples identified the same violations of the pattern which are opened by changing the tension of facial muscles, the opposite to the standart reference, or in the reaction of absence. The changing pattern of emotions, both in the first and in the second samples in the group of healthy test subjects is often marked on one lead. For the main group has been taken place frequent violations of the pattern on several leads. Also, they are characterized by the same type of distortion of the pattern in both samples according to the emotions of fear, sadness and anger. This phenomenon is called "chiralic effect". This effect allows to track the feedback performance both at the level of afferent synthesis, and at the level of the efferent link of SER.

The analysis of the experimental data, obtained on the basis of the facial feedback monitoring, opens the features of the SER feedback mechanism in epilepsy. The difference is due to the presence of the block, which manifests itself at all levels of feedback – the modulation of emotions (sample No. 1) and its simulation (sample № 2). In the main group the marked block is of a rigid character: there are systemic distortions of the facial pattern. The isolation as a special case of the block conserves the possibility of modulation of emotions when the one needed. At the same time it is absent an accidental and inadvertent activation of emotions. For example, an isolation of affect may to limit the effect of emotional resonance that occurs in natural situations of interaction with other people. Thus, the insulation contributes to the preservation stability without increasing the rigidity of PES. Blocking of the emotion is accompanied by the exception as of modulation as PES play associated with it. As shown by the analysis of the phenomenon of block at the level of the facial feedback associated with the emergence of the similar pattern of deviations from the feeling standard in the samples and is associated with the phenomenon of excessive rigidity. The identified signs of the block and insulation at the level of the facial feedback allow to separate the signs of collapse of emotion pattern from self-organizing processes. Comparison with the clinical data and results of quantitative stability estimates obtained on the basis of the matrix method, allows to confirm this hypothesis.

## 2. Analytical method of the stable qualitative estimation for the psycho-emotional state

As the basis for quantitative investigation of the psycho-emotional stable of patients with epilepsy the matrix method selected so it has well developed analytical apparatus having on the one side a sufficient ease of use and on the other hand wide use in accurate, social and humanitarian sciences [5].

Under developing the matrix of basal emotions (MBE) as the basis of the stability criterion assessment the following principles were used:

- the principle of the MBE symmetry (as a tool of the stability estimation for the psycho-emotional state);
- the principle of superposition (hypothesis) of the basal emotions – a possibility imposing) possibility to summarize (additively property) for the various emotions.

Besides, here is a subjectively – objective evidence of such a summarizing of the myograph readings for basal emotions.

So, for example, the rare person has not experienced at the same time a combination of such paired affects as joy and sadness, fear and angry, etc. At the same time such subjective sensations have an objective component: the statistic significant interview of psychic-healthy adults doubtless will prove this fact.

In addition the summation of the myograph readings for the paired affect analysis is confirmed due to that the summarized affect has the same dimension (mW) as separate affects.

The third specific scientific principle used in this research is the organizational unity and functional similarity principle which states: the facial reaction quality encoreded in facial profile that or another state must appropriate to intensive and extensive parameters of given emotions.

A mimic pattern of the basal emotions carries information about the state, simultaneously performing a regulatory function. This provision is consistent with the theory of basal systems of the emotional regulation (V. V. Lebedinsky, E. R. Baenskaya, M. M. Liebling, Nikolskaya O. S.), which is based on the research of early infantile autism. It is allocated the emotion property, aimed at achieving sustainability of the mental status due to its ability to regulate itself. So, an emotion as a functional system, and a system-forming factor which is the achievement and maintenance of a certain state is realized simultaneously in the format of an acceptor of result of action and in the one itself. Thus the feedback of emotion coordinates as the process of the PES changing in general so the authenticity of affective reactions. In this regard, the mental and physiological emotion reflect the individual links of the afferent and efferent synthesis at different stages of the evolution of PES, as well as afferent and efferent channels of the feedback of emotion. At the same time this mimic pattern of emotions is a specific form of manifestation and simultaneously coding of experience or state (V.V. Volov, 2007) [6]. This situation radically changes the usual interpretation of the adaptive effect of emotions in the immediate adaptation to external conditions, which mediates mimic and other physiological reactions.

When developing of the basal emotions matrix $\{Z_{ij}\}$ was used dyadic analysis of paired affects, selected on the basis of psychological polarity.

As noted above, in determining the matrix elements, we rely on the principle of superposition of the basic emotions. In this regard, in the general view of the matrix element of MBE has the following form:

$$Z_{ij} = Y_j + X_i, \; where \; i = 1 \div 3; \; j = 1 \div 3, \qquad 1.$$

but the matrix (MBE) has the following form:

$$Z_{ij} = \begin{Bmatrix} Z_{11} & Z_{12} & Z_{13} \\ Z_{21} & Z_{22} & Z_{23} \\ Z_{31} & Z_{32} & Z_{33} \end{Bmatrix}, \qquad 2.$$

where elements of this matrix are determined by adding the relevant basal emotions $(Y_i, X_j)$, as follows:

Table 1

|  | $X_1$ (angry) | $X_2$ (joy) | $X_3$ (surprise) |
|---|---|---|---|
| $Y_1$ (fear) | $Z_{11}$ | $Z_{12}$ | $Z_{13}$ |
| $Y_2$ (sadness) | $Z_{21}$ | $Z_{22}$ | $Z_{23}$ |
| $Y_3$ (disgust) | $Z_{31}$ | $Z_{32}$ | $Z_{33}$ |

For example, $Z_{11} = X_1 + Y_1$ is the summation of averaged electromyograph readings from the three leads of the average power extracted from the facial pattern leads of emotions of anger and fear. In a similar manner are prepared the elements of the matrix:

$Z_{22} = X_2 + Y_2 -$ is the summation of the testimony of the electromyograph accordings of the average tension values from the three leads on the emotions of joy and sorrow;

$Z_{33} = X_3 + Y_3$ – is the summation of testimony of the electromyograph accordings of the average tension values from the three leads on the emotions of surprise and disgust.

Off-diagonal elements of MBE are defined as follows:

$Z_{12} = Y_1 + X_2 -$ is the summation of testimony of the electromyograph accordings of the average tension values from the three leads on the emotions of fear and joy;

$Z_{21} = Y_2 + X_1 -$ is the summation of testimony of the electromyograph accordings of the average tension values from the three leads on the emotions of sadness and anger;

$Z_{13} = Y_1 + X_3$ – is the summation of testimony of the electromyograph accordings of the average tension values from three leads on the emotions of fear and surprise;

$Z_{31} = Y_3 + X_1$ – is the summation of testimony of the electromyograph accordings of the average tension values from the three leads on the emotions of disgust and anger;

$Z_{23} = Y_2 + X_3$ – is the summation of testimony of the electromyograph accordings of the average tension values from the three leads on the emotions of sorrow and surprise;

$Z_{32} = Y_3 + X_2$ – is the summation of testimony of the electromyograph accordings of the average tension values from the three leads on the emotions of disgust and joy.

It is known that the matrix criteria of stability were researched in [7]. In the present work is proposed three static stability criteria of the emotional state of the individual based on MBE.

The first is the deviations from the symmetry of MBE which are determined by the average value of deviations of the off-diagonal pair matrix elements from each other $(z_{ij}, z_{ji})$. Here as one of the principles for the design of a stability criterion is used the principle of symmetry, which in this case is transformed into symmetry of the proposed matrix, i.e. the off-diagonal pairwise elements of the one, ideally, should be identically equal to each other. The stability criterion is a measure of the deviation from symmetry of the matrix:

$$\langle \varepsilon \rangle = \sum_{i=1}^{3}\sum_{j=1}^{3} \mathrm{abs}(z_{ij} - z_{ji})/z_{ij(ji)}^{\max}/3, \qquad 3.$$

where $z_{ij(ji)}^{\max}$ is the maximum value of the off-diagonal MBE element.

It should be emphasized that the degree of deviation from the MBE symmetry of the patients with epilepsy should be compared with the corresponding values of MBE of mentally healthy balanced people.

The second static criterion is the value of tension of the psycho-emotional state (the matrix trace) or norm (power) of MBE, which is compared with the corresponding values of these criteria at the subsequent time (determined by the variation of tension values of MBE from the earlier states) or is compared with corresponding characteristics of a mentally healthy balanced people.

The trace of the matrix (MBE), representing the tension of the emotional state of the individual, is determined by summing of the EMG readings of six basic emotions:

$$L = Y_1 + Y_2 + Y_3 + X_1 + X_2 + X_3 \text{ (mW)} \qquad 3.$$

Criterion $\langle \varepsilon \rangle$ is the dimensionless value, and the mated with the one the dimensional value is $\langle \varepsilon * L \rangle$, which determines the share of energy of the PES feedback going on to the MBE imbalance.

In our case, all of MBE are canonical, and their elements are positively defined. This means that the norm of the matrix (or its power), calculated according to [5], as follows:

k – norm $$\|M_{ij}\| = \sqrt{\sum_{i=1}^{3}\sum_{j=1}^{3} z_{ij}^2} \,. \qquad 4$$

The one will be more of any elements of MBE:

$$\|M_{ij}\| \geq z_{ij}. \qquad 5$$

The third criterion is the indicator of the energy efficiency facial feedback of PES is defined as follows:

$$I = 1 - \left[ \sum_{i=1}^{3}\sum_{j=1}^{3} abs(z_{ij} - z_{ji}) / \|M_{ij}\| + \Delta L / \|M_{ij}\| \right] \cdot \varphi, \qquad 6.$$

where $\varphi = \begin{cases} 1, & \text{если } abs[\|M_{ij}\| - L)/\|M_{ij}\|] \leq \varepsilon_1; \\ \beta, & \text{если } abs[\|M_{ij}\| - L)/\|M_{ij}\|] > \varepsilon_1 \end{cases};$

$\beta = 0.7; \quad \varepsilon_1 = 0.05;$  $\qquad \Delta L = \sum_{i=1}^{3} abs(L/3 - z_{ii}).$

It should be noted that the energy of the facial feedback response is only part of the PES energy. Therefore, for the case when the MBE tension $L$ is nearest to value of norm of the matrix (or its power) the energy imbalance of MBE is gleaned from the emotional energy of the facial feedback ($\varphi = 1$) (6). Otherwise, share of the energy going on the MBE imbalance is transformed from the total PES energy ($\varphi = 0.7$).

As a graphic illustration of MBE is a diagram (Fig.1). It must be emphasized that for real quantitative measurements of MBE this diagram can only be displayed in a curved space.

$X_1$ – anger; $X_2$ – joy; $X_3$ – surprise; $Y_1$ – fear; $Y_2$ –sadness; $Y_3$ –disgust.

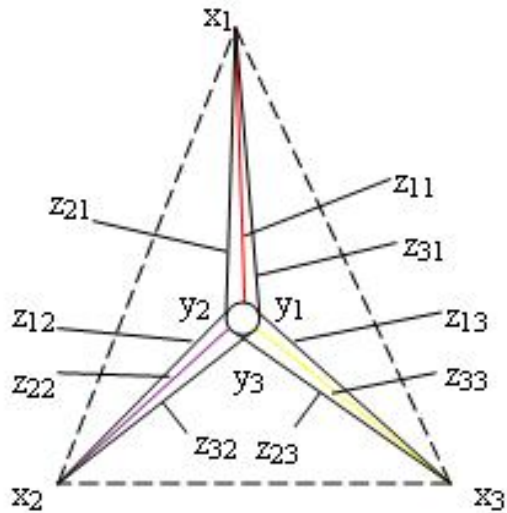

Fig. 1.

The analysis of the developed matrices of the psycho-emotional state of epileptics received at the EMG research result were realized in several stages. In the first stage, it is realized an integrative analysis of the psycho-emotional state of patients with epilepsy and healthy individuals based on the proposed energy criteria: norm (power), tension (trail) and the degree of energy imbalance, manifested in the deviation from the MBE symmetry. Table 2 presents the results of the comparative MBE analysis for healthy people and patients with epilepsy. The processing of the EMG data has revealed a trend: than higher the level of tension that is equivalent to an increase of the numerical value of the MBE tension (trace) (or power of a matrix), the higher the level of imbalance of the emotional state of the individual, which in turn is equivalent to the increase in the MBE asymmetry degree (Fig. 2-9).

Table 2

| The average value of the EMG indicators (healthy) | | The average value of the EMG indicators (epileptics) | | Differences in % | |
|---|---|---|---|---|---|
| $\langle L \rangle$ | $\langle \varepsilon \rangle$ | $\langle L \rangle$ | $\langle \varepsilon \rangle$ | $\langle L \rangle$ | $\langle \varepsilon \rangle$ |
| 51 | 14.1% | 62.3 | 19.2% | 20 | 26% |

The second stage consisted in the analysis of the constructed MBE matrices on the myographic research basis: it is determined the rank of matrices, especialities of degenerated matrices (it is classes of degeneration, determined the forms of interrelationships of basal emotions in MBE), the analysis of the MBE eigenvalues was realized. The analysis showed that all matrices are canonical and degenerate (rank of all matrices is equal two $r=2$). This fact says that the rows and columns of MBE are linearly-dependent, it tell us about the interdependence of paired affects of basal emotions of the individual. For example, under linear dependence of two rows or two columns in MBE it takes place the interdependence of five from the six basic emotions. In this stage is examining the types of relationship between paired affects of the psycho-emotional state of patients with epilepsy.

The third stage consisted in the comparison of the results of the selected groups analysis of patients with epilepsy at clinical and psychological characteristics with the results of the data group analysis based on the matrix method (section 4).

Table 3 presents the values of the PES energy imbalance of patients with epilepsy for the four classes of reactions. The first comparison of the analysis of epilepsy patients results based on the matrix method was carried out with the results of the clinical classification analysis, which showed their approval: all energy characteristics associated with the MBE imbalance (an indicator of the PES energy efficiency, the off-diagonal imbalance of the MBE elements, the PES intensity) correlate with their clinical assessments.

Analysis of the normalized MBE eigenvalues $(\lambda_1, \lambda_2, \lambda_3)$ in average showed, for example, that the average value $\langle \lambda_1 \rangle$ for patients with epilepsy differs by several orders of magnitude from the corresponding value in mentally healthy balanced people.

The average of values $\langle \lambda_2 \rangle$ for healthy persons and epileptics have the same value and approximately equal to one. The average value $\langle \lambda_3 \rangle$ of healthy persons corresponds to the maximum averaged value $\langle \lambda_3 \rangle$ of epilepsy patients allocated to class 3.

The question arises, what is the sense it has carried the analysis of the MBE eigenvalues and what are the magnitude of the MBE eigenvalues $\lambda_i$?

The first, it should be noted that $\lambda_i$ $(i = 1 \div 3)$ is some fraction of the energy facial feedback of PES enclosed in MBE, because $\lambda_i$ is measured in the same units of measure as the other elements of MBE (mW). If we multiply the normalized value $\lambda_i$ by the power of MBE, we get the absolute value of these shares of emotional energy of the facial feedback in mW. So, for example, the values of $\langle \lambda_1 \rangle$ and $\langle \lambda_3 \rangle$ are several orders of magnitude smaller than the averaged values of the norm (power) of MBE ($\langle \lambda_1 \rangle$, $\langle \lambda_3 \rangle \approx 0,001 \div 0,1 \quad mW$). Thus, in the analysis of MBE has identified that there is a hierarchy of energy levels: the first level is the power MBE - is energy of the facial feedback (corresponds to $\langle \lambda_2 \rangle$), the second level is the share of mental and emotional energy of facial feedback, going to the MBE imbalance, which is 10÷30% of the MBE power, i.e. it is the value of the order of 10 mW, and finally, the third level is the energy contained in the shares of the energy MBE eigenvalues $(\lambda_1, \lambda_2)$. As a hypothesis we can assume that the third energy level of MBE correlates with the energy of afferent and efferent synthesis of the facial feedback (P. K. Anokhin).

In section 4 we will give a detailed analysis of matching the quantitative research results of the facial feedback on the basis of the developed matrix method with the clinical and psychological researches of patients with epilepsy.

## 3. Entropy criterion of dynamic stability of psycho-emotional state of patients with epilepsy

In addition to the developed static stability criteria of PES, it is proposed criterion simultaneously performing the role of both static and dynamic criteria of MBE. As this criterion it is reasonable to use the conditional entropy H, which allows to use the analytical tools of synergetics. Stability theory developed in the works of A. M. Lyapunov [8] gave the theoretical-methodological instruments for the formulation of sustainability criteria in synergetics, and in greater detail in the thermodynamics of the structure, where the sustainability criteria adopted marks of the second differential of entropy (close to an equilibrium) and the minimum of the excess entropy production for system states are placed far from equilibrium [9]. In this regard, as a general measure of psycho-emotional state and the basis of sustainability criteria of the individual has developed the conditional entropy formula as follows:

$$H = \alpha\left[1 - \delta(M_{ij}^{(sym)} - M_{ij})\right] + \frac{H_0 - \alpha}{\ln(a^{r-1})}\ln\left[a^{r-1} + b\left(\frac{\widetilde{\varepsilon}L}{\langle\langle\widetilde{\varepsilon}L\rangle_{healthy}\rangle} + \frac{\Delta\overline{L}}{\langle\Delta\overline{L}_{healthy}\rangle} + \frac{\Delta\overline{E}}{\langle\Delta\overline{E}_{healthy}\rangle} - 3\right)\right]\left[1 - \delta(r-1)\right], \quad 3.$$

In formula (3) $M_{ij}^{sim}$ is the symmetric of MBE, with the same norm as MBE, which received in the myographic research result.

$$\alpha = 0.3, \quad a = 1.2, \quad b = 1\cdot 10^{-3}, \quad \delta(M_{ij}^{sim} - M_{ij}) = \begin{cases} 1, & M_{ij}^{(sim)} = M_{ij} \\ 0, & M_{ij}^{(sim)} \neq M_{ij} \end{cases}, \quad \delta(r-1) = \begin{cases} 1, & r = 1 \\ 0, & r \neq 1 \end{cases}, \quad 4.$$

$$\text{где } \overline{\Delta E} = \sum_{i=1}^{3}\sum_{j=1}^{3} abs(z_{ij} - z_{ji})/L; \quad \Delta\overline{L} = \langle\Delta L/L\rangle_{healthy} = 0.1023;$$

$$\langle\overline{\Delta E}\rangle_{healthy} = \langle\Delta E/L\rangle_{healthy} = 0.2047.$$

In the formula (4) $r$ – is a rank of MBE, $\langle\widetilde{\varepsilon}L\rangle$ is the value of the averaged energy MBE imbalance of the individual, $H_0 = 0.618$, is the value of entropy of the "Golden section" $\langle\widetilde{\varepsilon}L\rangle_{healthy}$, $\langle\Delta\overline{E}\rangle_{healthy}$, $\langle\Delta\overline{L}\rangle_{healthy}$ – the averaged values of the energy imbalance, relative of MBE imbalance the off-diagonal elements and the heterogeneity of tension for a healthy balanced people respectively.

Conditional entropy of MBE $H$ meets all rules of the mathematical entropy [9]:

1) $H \geq 0$ is positivity of entropy;

2) $H \subset (0;1)$;

3) $H = \sum_{i=1}^{n} H_i$ is the property of additivity.

Fig. 2 shows dynamics of the conditional entropy of the emotional state based on the EMG researches of patients with epilepsy. From this picture it follows that despite the different initial level of entropy ($H = 0.576$ to A. M. A. the patient and $H = 0.637$ for the patient A. A. E.) at the stage of evolution of both patients the PES entropy approaches to the "Golden section", i.e. practically to normal psycho-emotional state. Each sequence group number corresponding to a larger value of entropy can be mapped with some native time functioning of the psyche of the averaged individual for the group. Fig. 3 shows the values of the one of healthy and epileptics with increasing of the integrative measure $H$.

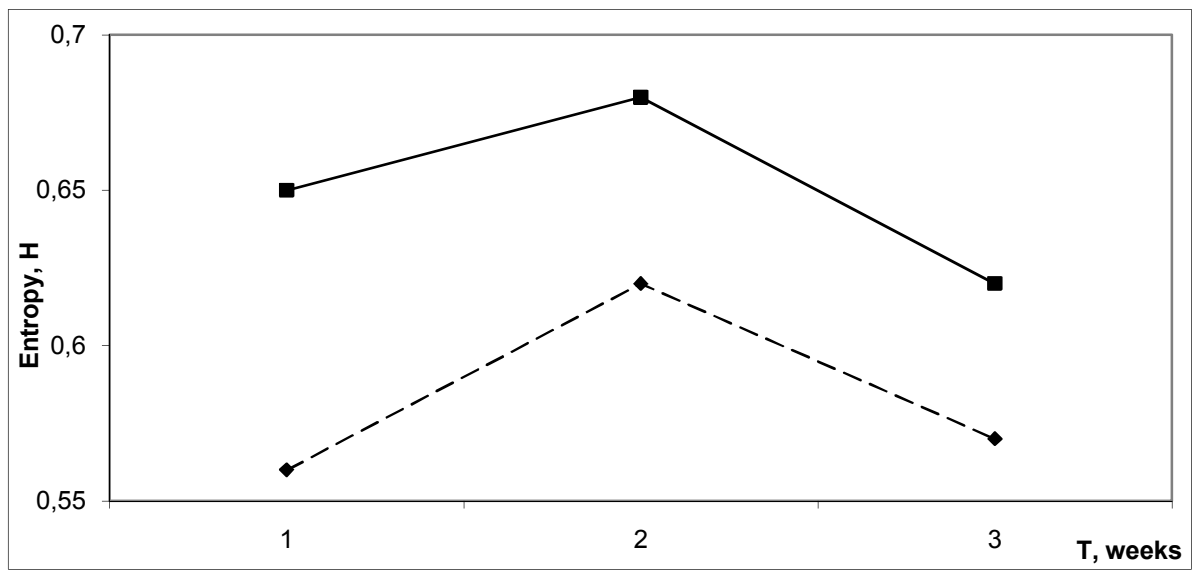

Fig. 2 Dependence of the conditional MBE entropy from the time for epileptics

▲ - the first patient E; ♦ - the second patient M.

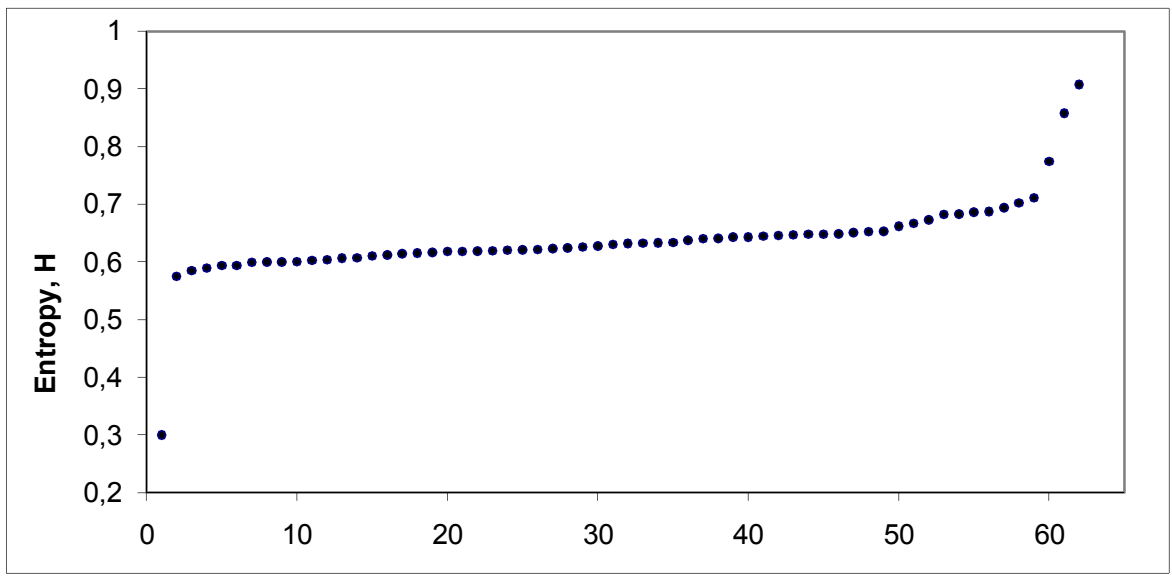

Fig. 3 Dependence of PES entropy from the number of the patient with increasing the entropy value

On the basal emotion matrix basis has been formulated its conditional entropy, which represents an integrative measure of the variation of the energy PES imbalance of the individual with the corresponding averaged value of the energy PES imbalance of mentally healthy balanced persons. This criterion allows to use the analytical apparatus of synergetics to analyze the static and dynamic stability of psycho-emotional state of the individual. The proposed conditional integrative measure for the emotional state assessment of the individual - entropy $H$ – it allows to give additional information about static and dynamic emotional state of the individual and the patient with epilepsy.

Developed qualitative and quantitative instruments of the research of the psycho-emotional state of patients with epilepsy allows to realized an integrative analysis of the EMG researches results.

## 4. Integrative analysis of emotional state on the basis of quantitative and qualitative research methods

In this section it is realized the analyses of comparison of the quantitative and qualitative results of myographic investigations of facial reactions, which reflect together the features of PES of patients with epilepsy. Measuring of the amplitude-frequency indicators of muscle tone in the process of emotional response presents energy characteristics to assess the condition of the basal emotions matrix (MBE), its stability, obtained by using of the proposed matrix method. The facial response detection in the form of deviations from the reference patterns of emotions in the experimental samples is the basis of estimating the facial feedback mechanisms in SER.

The definition of the nature of the reaction in a single complex of the mimic pattern (multiple leads) reflects the work of facial feedback of emotions in it the qualitative content (S. Tomkins, E. Gellhorn): any deviations from the standard have certain grounds. A special attention was paid to chiralic effects and the phenomenon of blocking (facial reactions, naturally diverging with the standard), the definition of which formed the basis of the model of quality diagnostics of basal emotions.

The comparison of results of the quantitative and qualitative analysis according to EMG researches have allowed to identify the features of the facial feedback mechanism of basal emotions among patients with epilepsy and to determine the state of their psychological and emotional sphere. It is this innovative approach, based on model of quality diagnostics of basal emotions and developed the matrix method for quantitative evaluation, including the generalized criteria of efficiency of the emotional sphere functioning of the individual, allows to estimate, the tension, the conditional entropy of the system "mentality" within the experimental samples.

In this paper the research of facial reactions are considered as of a feedback profile in the system of emotional response (SER). This provision is new in presention of this phenomenon of facial expressions – not only as non-verbal accompaniment of emotions on a physical level (i.e., fixating on a motor level shifts in the sensuous sphere), but primarily as a physiological trigger of the feedback functional mental systems of the self-organization.

The research involved patients with epilepsy with different origins and clinical forms. All of the subjects are different in that intellectually intact (no signs of dementia), there is no pronounced morphology (brain substrate in the form of cysts, tumors, etc.). The selection of the main sample realizes in view of clinical factors, but not based on it, and it comes from the main idea of the work – determining of the self-organization attributes and collapse signs of the mimic pattern as the indicator of MBE. The comparison group included healthy individuals without neurological disorders. The method has been calibrated by using this group. First of all, were normalized EMG samples and assessments of key indicators. There was made the comparison of data on developed parameters by using the qualitative diagnosis model of basal emotions (myographic scheme and evaluation criteria) and on indicators of the matrix method (tension of MBE, the level of imbalance, the indicator of energy efficiency, conditional entropy, etc.).

Given investigations were analyzed from several positions. In the present work, the first test captures emotional reactions, with arbitrarily occur at the subject. The second test is organized in an experimental organized situation of an involuntary emotional regulation: when the subject is offered to identify the emotion in the stimulus material (shown in the photo), his face registered facial reactions.

In the first case an ability for meaningful inclusion is fixed in one or another emotional state (PES) with the assessment of the quality of the facial feedback (compliance to the standard, deviation, blocks). In the second case the triggering of the so-called effect of emotional resonance associated with the archaic offline functional system of mimic reflection estimated. A comparison between these two samples on the same emotion allows on the level of qualitative MBE assessments to diagnose the condition of SER, its stability and the safety of the fundamental mechanism – the feedback. The analysis carried out on the basis of the matrix method, presents quantitative characteristics of the MBE tension and energy costs related to its sustainability.

Previously in the research of the power recording of EMG it was conducted a quantitative analysis of paired affects and non-polar basic emotions (fear, anger and sadness) in patients with epilepsy and healthy persons. It is with these emotions is associated many symptoms accompanied the epilepsy clinic. The analysis showed the following: in the group of healthy men anger prevails over fear in almost 100% of cases. Observed women shows the opposite pattern: in 62.5% of women the maximum tension of facial muscles is recorded by on the emotion of fear. The same trend is observed in patients with epilepsy: in men anger prevails over fear in 60% of cases, and in women, by contrast, fear prevails over anger at 63.6% of cases. These results characterize the gender aspect of PES, i.e. in women irrespective of the illness, the tension detected at the level of the facial feedback regardless of disease higher on the emotion of fear than anger in a uniform proportion of cases.

Comparing the data of quantitative analysis of the non-polar basal emotions of fear and sadness in the framework of MBE method, also obtained certain trends. For the power recording parameter of EMG in the group of healthy men sadness prevails over fear in 80% of cases. In 60% women fear prevails over sadness. In the group of patients with epilepsy (men) the PES tension is higher fear in 57,2% of cases, and in women fear prevails sadness in 94% of cases. These results on the energy level reflect the specifics of PES of epilepsy patients.

Additionally, it was made a comparative analysis of the results on the energy balance with the character of quality of the reaction. So, in women with epilepsy on the emotion of fear revealed the antagonistic block in excess of the PES tension in comparison with sorrow. Also they show the predominance of total block at a prevalence of fear over anger. In the cases of the energy advantage of anger over fear are marked characteristic reactions of violation for the pattern emotion of anger. Thus, women of a core group in cases of an energy advantage in the emotion of fear is placed of blocking this affect. A similar trend is observed and the emotion of anger (in the case of the prevalence of anger over fear), demonstrates the mechanisms of stability associated with an imbalance of paired and non-paired basal emotions, at the level of the facial feedback. In men the main group at a prevalence of fear observed all the forms of the block to this affect. However, as in the control group, any trends not revealed too.

For integrative analysis of the PES individual on the basis of a comparison of quantitative data obtained by using the matrix method, with the results of the pilot research, identified by the model quality diagnostics of basal emotions, the subjects of the core sample were divided into several groups.

The clinical principle formed the basis of the first classification. The application of new evaluation criteria for energy performance of MBE allowed to reveal the individual profile of the state of psycho-emotional sphere, determine its stability. In turn, the comparison of data on leading indicators of the matrix method an (indicator of energy efficiency associated with energy consumption for the imbalance, PES tension, the MBE entropy) groups with different levels of pathology allowed us to determine the validity of the selected model. The results of the matrix analysis have presented in table 3.

Table 3

| Energy performances of various clinical groups | Rare generalized seizures (group 1) | Frequent generalized seizures (group 2) | Partial seizures equivalents (group 3) | Remission (group 4) |
|---|---|---|---|---|
| $\langle \varepsilon \rangle$ | 0.18 | 0.23 | 0.22 | 0.17 |
| $\langle L \rangle$ | 49.6 | 85.86 | 61.77 | 77.71 |
| $\langle \varepsilon L \rangle$ | 9.59 | 20.5 | 14.6 | 14.4 |
| $\langle H \rangle$ | 0.613 | 0.66 | 0.64 | 0.64 |
| $\langle I \rangle$ | 0.67 | 0.45 | 0.73 | 0.71 |

Analysis of the first classification showed a number of fundamental differences among clinical groups of subjects. In particular, indicators of group 2 differed from group 1 with low energy efficiency $\langle I \rangle$, a higher tension $\langle L \rangle$, large amounts of energy imbalance on $\langle \varepsilon L \rangle$ and higher level of the MBE entropy $\langle H \rangle$. Interesting data were obtained in the group of epileptic equivalents (3) and the group with absence paroxysms in the clinic from one to three years or more (4). Both groups possess a high energy efficiency of PES compared with groups with generalized paroxysms, as well as the best rates by other criteria in comparison with the group with frequent attacks. However, the comparative analysis of these group data with data of the group 1 (rare generalized paroxysms) showed mixed results. Thus, indicators for tension, the energy consumption for the imbalance and MBE entropy was slightly better in the first group. This is not to say contradictions of the results obtained by using the matrix method, but rather adds to questions about the diagnosis of PES on sustainability factors of MBE. The above features of the first group distinguished it from the rest, partly was grounded in the concept of A. G. Ivanov-Smolensky, according to which convulsions, being archaic defensive reaction, contribute to the discharge of paroxysmal brain and the release of toxins by the acute metabolic adjustment.

A kind of "reboot" triggs. This explains the somewhat lower values of tension and level of imbalance of MBE in the group with occasional larger attacks. However, on the main index – energy-efficiency – results are significantly better in groups without generalized scamps in the clinic, which corresponds to the more effective clinical picture. This generalization is consistent with the position of the energy paradigm.

Thus, the comparison of the results of quantitative estimates of the MBE stability obtained on the basis of the matrix method, clinical data analysis, showed their compliance. This is confirmed in the evaluation of the energy characteristics of the additional group of epilepsy patients with intact higher mental functions, differing in speech, thinking, speed of the associative process, capacity and social activity with other groups. The parameters of the MBE in this group significantly exceeded those of the group with frequent generalized convulsions. For all groups this values are close. Attracts the attention of a sufficiently high tension level $(L = 65.3)$ with low imbalance $(\varepsilon = 0.179)$.

Separately, a comparative analysis of the subjects, able to predict the impending paroxysm and patients that do not have this feature was realized. The last group was selected to determine qualitative and quantitative differences in the energy parameters and the static stability criteria associated with signs of natural self-organization [1]. Comparative data are shown in table 4.

Table 4

| Energy perfomances | Groups with harbingers | Groups without harbingers |
|---|---|---|
| $\langle \varepsilon \rangle$ | 0.17 | 0.2 |
| $\langle L \rangle$ | 57.5 | 66.19 |
| $\langle \varepsilon L \rangle$ | 9.99 | 15.2 |
| $\langle H \rangle$ | 0.6 | 0.65 |
| $\langle I \rangle$ | 0.7 | 0.6 |

It is interesting to note that in the group of subjects that has function for probabilistic forecasting (V. Simonov) has revealed the highest values of energy efficiency $\langle I \rangle$. Foresight and in some cases, the ability to suspend the attack characterizes this group of low values for energy imbalance costs and the lowest MBE entropy, almost close to the reference values of healthy, among all of the groups identified by the clinical principle $\langle H \rangle$.

The principle of chirality was laid in the second group of classifications. According to this principle, the subjects were divided into (on) qualitative reactions observed simultaneously in the first and second experimental samples. First of all, were allocated chiralic (mirror) reactions associated with the violation of a reference expression pattern of the individual basal emotions. In the work this reaction indicates as a manifestation of the so-called block. This phenomenon is described in literature as fixation of affect, and jam of the affect, etc. In the researches of clinical psychologists this phenomenon is associated with alexithymia, psychosomatics, conversion syndrome, a psychological manifestation protective mechanisms and a number of other normal and pathological phenomena.

In the process of experimental samples it identifies the following forms of chiralic effects: 1. the antagonistic block; 2. the partially distorted pattern; 3. the total unit; 4. the asymmetric block.

The second classification, built on the principle of chirality, where selected - two groups of subjects. The idea was to compare typical reactions playable on one emotion in two samples. Spotted trend was the basis for the classification.

Reactions were considered separately with obvious signs of block emotions (the antagonistic reactions and the total block), as well as profiles of reactions without any effects. The separation on the dyad allowed to determine the effectiveness of compared forms of response associated with chiralic effect. The results of the matrix analysis presentes in table 5.

Table 5

| Energy indicators groups with different chiralic effects | 1 pair of comparison | | 2 pair of comparison | | 3 pair of comparison | | 4 pair of comparison | |
|---|---|---|---|---|---|---|---|---|
| | group (1-1) | group (1-2) | group (2-1) | group (2-2) | group (3-1) | group (3-2) | group (4-1) | group (4-2) |
| $\langle \varepsilon \rangle$ | 0.21 | 0.22 | 0.189 | 0.197 | 0.228 | 0.187 | 0.205 | 0.193 |
| $\langle L \rangle$ | 65.92 | 60.04 | 51.23 | 60.28 | 50.29 | 63.5 | 69.97 | 56.85 |
| $\langle \varepsilon L \rangle$ | 14.30 | 14.93 | 10.004 | 12.857 | 12.37 | 12.727 | 14.87 | 11.83 |
| $\langle H \rangle$ | 0.646 | 0.638 | 0.624 | 0.632 | 0.639 | 0.63 | 0.642 | 0.620 |
| $\langle I \rangle$ | 0.606 | 0.638 | 0.673 | 0.665 | 0.603 | 0.681 | 0.636 | 0.636 |

In the first pair of comparison included patients with the presence of antagonistic reactions (1-1), noted in both the first and second samples on the first emotion, and in the other patients with partially impaired pattern of basal emotions (1-2), also repeated in both samples. The differences were insignificant. The results of energy consumption for the imbalance and the level of conditional entropy $\langle H \rangle$ are better in the second group. The tension level is slightly higher in the first group (against 60.04). Indicators of the integral criterion is an indicator of the energy efficiency (IEE) – are better in the second subgroup. However, it is insignificant (I=0.638, I=0.606).

Thus, both groups have close energy characteristics of MBE. Apparently the insignificance of differences between groups connects to the fact that regardless of the block form (whole the one with respect to the standart reaction pattern of emotions or partial), the only one task is solved under similar conditions. The second pair consisted of patients with the presence of the partially broken pattern or reaction with the correct pattern (2-1), which is mirror reproded in both samples;

the pair group (2-2) consists the cases with asymmetric reaction (profile emotions, regardless of conformity to the standard, it changes to the opposite trends in the second sample). This couple also doesn`t reveal significant differences, primarily according to the criteria of $\langle I \rangle$ and $\langle H \rangle$. However the group 2-2 distinguishes the values of the criteria of tension PES $\langle L \rangle$ and energy imbalance of MBE $\langle \varepsilon L \rangle$, although marginally. Preliminary conclusion: the differences in the decision of the regulation form result in both groups is achieved the same one. Asymmetric reactions are detected as an exclusive profile of the facial response of the experimental SER model. In this case it says that for this group 4 needs a special solution to achieve static stability in terms of excessive rigidity. The group 2-1 (and, hence, also of the group 1-1 and 1-2) solves the problem of the sustainability achieving of MBE, based on return conditions – is a lack of sustainability. Regardless of differences in the profile of the block of the last format, characterizes the bulk of the basic subjects of the basic sample. According to the clinical data asymmetric reactions occur often in the first group, group less in the second group, i.e. in groups with large paroxysms. It means that in terms of total threat of the psychopathology resistance loss is formed by the excessive rigidity of MBE, offset, in turn, due to the asymmetric profile of SER.

The third pair classification is built on the basis of the presence of antagonistic block in the first (3-1) or in the second sample (3-2). Accordingly, these groups are identified as groups with obvious or the hidden block. According to the results of the matrix method using were defined indicators and the MBE criteria for two groups. The values of the energy efficiency indicator of PES $\langle I \rangle$ are better for the group 3-2. This fundamentally distinguishes it from the group 3-1, it is said testifying about the greater effectiveness of the hidden block – it is the limit of feedback on the level of the PES playing. In this group better performances are occurred on the level of conditional entropy $\langle H \rangle$. However, the higher indicator values on the trail $\langle L \rangle$ characterizes the MBE group 3-2 as more intense. Nevertheless, the level of the MBE imbalance is lower than in the group 3-1.

Preliminary conclusion: the both blocks are included in SER at the different levels, in different ways to solve the problem of stability preservation. Higher tension in the group with the hidden block is not accompanied by high level of the energy MBE imbalance, that qualitatively differs it from the other groups, where this given regularity is observed: the formation of antagonistic block in the first sample, the appearance of the chiralic effects in these conditions occurs as a factor in of the imbalance MBE leveling. Obviously, in cases of the so-called hidden block, it is solved another problem – is the preservation of stability: an antagonistic reaction in a situation of emotional resonance restricts involuntary involvement in something other state of PES (so, for example, it is in the process of the social interaction and communication with other people). The hidden block is mainly rare observed in the clinic in the group of subjects with frequent generalised convulsions. This trend is an example of the private existence of insulation emotions, which shows the energy advantage over the block is the alternative limitation feedback as a mechanism of stability.

The fourth pair of comparison was included the patients with the presence of the so-called total block (4-1) is the mostly reactions of increasing, more seldom reduction of tone on all samples and on all leads. In another group were selected the patients without signs of the block, with mixed reactions (4-2). The results of the group 4-2 on a number of indicators were better: the levels of the MBE energy imbalance $\langle \varepsilon L \rangle$, trail $\langle L \rangle$ were of the less it characterizes the PES 8 group as less tense. However, the values on the energy efficiency indicator (IEE) two groups $\langle I \rangle$, to esteem often equal that means the they are equally effective. As shown by the data analysis of the clinical profile, the most often total block occurs in the group of patients in remission and in patients with partial convulsions and mental equivalents. This is because the total block at its base, is an antagonistic reaction to fear, reflected in all MBE, and, therefore, with the affect associated risk of loss of emotional stability in these groups. Not by chance this group consists of representatives of clinically severe groups, and groups that are in remission. An additional analysis of the quantitative data of MBE obtained by using the matrix method reveals high prevalence of stress in this affect. The clinical data also reveal symptoms of anxiety in the precursors, either in the frameworks of the symptoms observed in megalocephaly period (between two paroxysms).

Summarizing the results of comparing two samples, we can conclude the following. The presence of type profile of emotions in two complementary samples that involves different mechanisms of facial feedback, demonstrates itself certain regularity inherent in epilepsy patients. Mirror effects of MBE (marked in 2-1, 2-2 gr.) demonstrate different forms of the static stability maintaining. Specific distortion of the emotion pattern, in our opinion, indicates the manifestation of the sustainable pathological state achieving mechanism (N. P. Bekhtereva), when due to the facial feedback limitations is achieved the temporary balance while maintaining the feedback as a leading mechanism of the functional system. A variety of the chiralic effect forms detected at the level of the facial response (1-1, 1-2, 2-1, 2-2 C.), describes the mental mechanisms of self-organization in conditions of paroxysmal brain, characterize variations of solving the problem of sustainability achieving.

In fact, the format for the two samples, allows to perform monitoring of the facial feedback: in the position of the entrance to the emotional state with the reflection on the face (the first test) and when it transfers to the one the reflection of facial reactions of the perceived emotions as the SER output (the second sample). In the first case, at the level of facial expressions is fixed the profile of the emotional response (pattern emotions, blocking, etc.). In the second case, the playback of emotions happens indirectly – starting from the facial feedback. Imposed in the experimental model in a single matrix the qualitative data analysis, the results of the entrance to the emotional state and the inclusion in the reverse direction through the conditional output (involuntary reproduction of facial reactions) show the palette of the feedback block.

In general view the experimental model allows to show of SER at the level of the feedback mechanism, as under the infancy of emotional state, and if reproducing due to the effect of the emotional resonance. Comparing the experimental results with the founds of P. K. Anokhin theory, one can identify the results of the first and second samples as indicators respectively of the efferent and afferent links of the MBE feedback within the framework of the functional system of PES. On the basis of integrative qualitative and quantitative analysis of MBE, hold on the results of the facial feedback research, the characteristic shapes of the SER feedback block of patients with epilepsy have been revealed.

So, the reconstruction in the experimental conditions of SER as a functional system represents the especialties of its organization and defects. The analysis conducted by using the specially developed matrix method, revealed quantitative characteristics of the MBE stability. In particular the energy profile shows the efficiency of the system functioning, the level of tension and the MBE imbalance, spendings on compensation, of the one and the level of it's the conditional entropy. This approach enables to estimate of PES, its energy capacity and stability the quantitative and qualitative level. Further comparison of these data with the results of a qualitative analysis, based on the principle of chirality, allowed us to determine the effectiveness of different forms of the feedback block as a specific mechanism to achieve the sustainability of PES. To conclude, the one the developed demonstration model of estimation of feedback mechanism within the framework of emotional response as functional system of diagnosis at different stages of its evolution. Inspite of the fact that myographic measurements to be made experimentally, reflect an instant snapshot of the system state that can change over time, obtained data, however, convey a picture of the device of functional mental systems of self-organization.

## Conclusion

In the article it has been developed methodology the research of the emotion feedback, including the principles of symmetry and superposition, the matrix of the basal emotions, mathematical apparatus of the theory of matrices, the model of quality diagnostics and program investigation based on the EMG measurements of the facial reactions. For the first time the qualitative-quantitative method of evaluating of PES of patients with epilepsy based on the study of facial feedback has been approbated. The tools of the quantitative assessment of PES of epileptics on the MBE basis including static and dynamic criteria has been carried out.

The research established revealed a trend: on average, with increasing power of the matrix of basal emotions (or the digital values of tension) increases its energy imbalance, and, therefore, the emotional state stability becomes less. The integrative analysis showed that the higher tension corresponds with higher performances of the energy imbalance. Identified on the basis of the qualitative analysis of experimental data research the antagonistic block on separate emotions in such circumstances is revealed as a leveling factor of the MBE imbalance.

A significant result of this research, obtained on the MBE method basis, is natural-mathematical proof that the paired basal effects are mutually dependent, because all MBE are strictly mathematically degenerate. For example, one from the possible cases degenerated matrix implies linear dependence of the rows and columns of MBE. However, only five of six basal emotions are linearly dependent. This actually substantiates theoretical assumptions about the unity of the psycho-emotional sphere, the relationship of basal emotions as functional systems.

The principal result of integrative analysis is that the comparison of data on qualitative and quantitative research methods of psychoemotional state of patients with epilepsy showed their congregation. Selected on the clinical signs four isolated groups, distinguished by the type of disease, shape and frequency of the paroxysm, reveal the energy differences that allows to identify these groups as clusters on coordinates of tension and imbalance in the structure of the psycho-emotional state of patients with epilepsy.

On the basal emotions matrix basis it has been formulated the conditional entropy, which represents an integrative measure of the variation of the energy PES imbalance of epileptics with the corresponding averaged value of this indicator in mentally healthy people. The proposed conditional integrative measure of the PES – the entropy ($H$) – allows to open additional information on the dynamics of the emotional state of the patient with epilepsy.

Returning to the methodological issue of determining the belonging of an object of the present research, we have to turn to theoretical positions and models of functional systems. In the theory of academician P. K. Anokhin any system organized for the specified result. This is determined by its procedural orientation and functional organization. A key element of this system is feedback, which, on the one hand, reports about the necessary changes associated with the outcome of the action, and, on the other hand, at the same time testifies to the compliance of the system itself. Considering the experimental data on PES of two samples this research, as a manifestation of the SER patients with epilepsy identified blocks and signs of isolation of emotions associated with the emergence of chiralic effects at the level of the facial feedback represent new mechanisms to achieve the result of the action due to the static stability.

## Acknowledgements

The authors are grateful to PhD., associate professor A. P. Zubarev for the help in the software processing of the experimental data.